\documentclass[pra,aps,twocolumn,amsmath,footinbib]{revtex4-2}
\usepackage{amssymb}
\usepackage{graphicx}
\usepackage{hyperref}

\usepackage{color} 

\definecolor{darkgreen}{rgb}{0,0.7,0}  

\usepackage[normalem]{ulem} 

\usepackage[table]{xcolor}
\definecolor{lightgray}{gray}{0.9}

\usepackage{booktabs}

\usepackage[caption=false]{subfig}
\newcommand{\phantomsubfloat}[1]{
    {
        \captionsetup[subfigure]{labelformat=empty}
        \subfloat[][]{#1}
    }%
}

\begin{document}

\title{The Whitham approach to Generalized Hydrodynamics}

\author{Frederik M{\o}ller\textsuperscript{1},
Philipp Sch{\"u}ttelkopf\textsuperscript{1},
J{\"o}rg Schmiedmayer\textsuperscript{1},
and Sebastian Erne\textsuperscript{1}
}

\affiliation{
\textsuperscript{ 1} Vienna Center for Quantum Science and Technology (VCQ), Atominstitut, TU Wien, Vienna, Austria\\
}

\date{\today}

\begin{abstract} 
The formation of dispersive shock waves in the one-dimensional Bose gas represents a limitation of Generalized Hydrodynamics (GHD) due to the coarse-grained nature of the theory.
Nevertheless, GHD accurately captures the long wavelength behavior indicating an implicit knowledge of the underlying microscopic physics.
Such representation are already known through the Whitham modulation theory, where dispersion-less equations describe the evolution of the slowly varying shock wave parameters.
Here we study the correspondence between Whithams approach to the Gross-Pitaevskii equation and GHD in the semi-classical limit.
Our findings enable the recovery of the shock wave solution directly from GHD simulations, which we demonstrate for both zero and finite temperature.
Additionally, we study how free expansion protocols affect the shock wave density and their implications for experimental detection.
The combined picture of Whitham and GHD lends itself to additional physical interpretation regarding the formation of shock waves.
Further, this picture exhibits clear analogies to the theory of Quantum GHD, and we discuss possible routes to establish an explicit connection between them. 
\end{abstract} 

\maketitle 

\section{Introduction}

Owing to experimental advances over the last few decades, gases of ultracold atoms have become a ubiquitus platform for studying out-of-equilibrium dynamics of many-body quantum systems~\cite{RevModPhys.80.885}.
Upon confinement to one dimension, a repulsively interacting Bose gas is well-described by the quantum integrable Lieb-Liniger model~\cite{lieb1963exact, LL2}, whose exact equilibrium solution, expressed in the form of collective quasi-particles, can be obtained by means of the Bethe Ansatz~\cite{Bethe1931}.
Out of equilibrium, quantum integrable systems exhibit very different behaviour compared to their higher-dimensional counterparts; owing to the presence of an infinite number of conservation laws, their dynamics are highly constrained and thermalization is inhibited~\cite{Gogolin2016}.

Recently, the theory of \textit{Generalized Hydrodynamics} (GHD)~\cite{castro2016emergent, bertini2016transport} was developed, applying general principles of hydrodynamics to quantum integrable systems:
By expressing the currents of conserved quantities in the basis of Bethe Ansatz quasi-particles, GHD achieves a dispersion-less, coarse-grained hydrodynamic description~\cite{Bulchandani_2017}.
Despite the conceived fragility of quantum integrability, GHD has successfully described dynamics of experimental systems far from equilibrium~\cite{schemmer2019generalized, malvania2020generalized, PhysRevLett.126.090602, PhysRevX.12.041032}.

However, the description provided by GHD also has its limitations; for instance, as a local density perturbation of a condensate expands, its density profile will eventually acquire a near infinite gradient, heralding a breaking of the wave. 
In a weakly interacting condensate, such an un-physical hydrodynamic gradient catastrophe is avoided by transferring energy from large to small length scales by the means of dispersion.
The result is a \textit{dispersive shock wave} (DSW) characterized by a nonlinear, oscillatory wave train of short wavelength, whose front propagates faster than the local sound velocity of the medium~\cite{kamchatnov2000nonlinear}.
The formation of DSWs has been observed in several experiments with ultracold atoms~\cite{doi:10.1126/science.1062527, PhysRevA.74.023623, PhysRevLett.101.170404, PhysRevA.80.043606} and fluids with light~\cite{PhysRevLett.99.043903, PhysRevX.4.021022, PhysRevLett.118.254101, PhysRevLett.126.183901}.
Such behavior of the condensate can be derived from the mean-field description provided by the \textit{Gross-Pitaevskii equation} (GPE), whose quantum pressure term represents the dispersive properties of the gas.
Meanwhile, GHD fails to capture the DSW, as its hydrodynamic equations are dispersion-less and its coarse-grained scale is much larger than the length scale of the DSW oscillations.
Nevertheless, the theory seemingly predicts mean quantities of the oscillating wave~\cite{https://doi.org/10.48550/arxiv.2208.06614}.

In classical fluid dynamics, the nonlinear dynamics of dispersive shock waves can be asymptotically represented through Whitham modulation theory~\cite{whitham2011linear}. 
By averaging a number of conservation laws of the underlying dispersive equation, here the GPE, over the period of the DSW, one obtains a system of dispersion-less hydrodynamic equations describing the slow modulation of the DSW amplitude, wavelength, and mean on a scale much greater than the travelling wave.
Conceptually, this separation of scales in Whitham's approach to classical systems is very similar to the assumptions of local equilibrium in the GHD approach to quantum system; 
indeed, the steady modulated solutions of the classical system are analogous to the locally stationary quantum states.
This remarkable observation was first pointed out in Ref.~\cite{Bettelheim_2020}, where an explicit correspondence between Whitham's approach and GHD in the semi-classical limit of the one-dimensional Bose gas was proven.
In separate numerical studies of GHD~\cite{PhysRevResearch.3.013098, https://doi.org/10.48550/arxiv.2208.06614}, suggestions of this similarity were also made. 
Further correspondence between the two approaches was established in Ref.~\cite{Bonnemain_2022}, where a GHD description of the soliton gas for the Korteweg–de Vries (KdV) equation was constructed.

In this work, the connection between Whitham's approach and Generalized Hydrodynamics is further explored.
We begin by briefly reviewing the zero-temperature GHD of the Lieb-Liniger model and the Gross-Pitaevskii equation (Sec.~\ref{sec:theory}).
Next, in Sec.~\ref{sec:Whitham}, we expand upon the results of Ref.~\cite{Bettelheim_2020} by explicitly relating the Riemann invariants of GPE Whitham theory to the Fermi rapidities of GHD and studying the convergence of the theories as one approaches the semi-classical limit.
Here we find a high correspondence between the theories in a regime accessible by current experiments.
In Sec.~\ref{sec:Riemann_problem} we demonstrate how Whitham's approach can be employed to recover the oscillating density of a dispersive shock wave from GHD simulations, where it would otherwise be inaccessible.
To this end we simulate dynamics of the bipartition problem at both zero and finite temperature.
Additionally, we study the evolution of a shock waves following free expansion and how such protocols may improve their experimental detection.
Section~\ref{sec:discussion} is devoted to discussing the physical interpretation of DSW formation and the generalization of Whitham's approach to arbitrary interaction strengths of the Bose gas.
We review the different nature of shock wave oscillations in the weakly and strongly interacting regimes and compare different approaches for describing said oscillations.
In particular, we conjecture that Whitham's theory of the 1D Bose gas might be the semi-classical limit of the Quantum Generalized Hydrodynamics~\cite{PhysRevLett.124.140603} and outline how one may proceed to establish a connection between the two approaches.
Finally, we conclude in Sec.~\ref{sec:conclusion}.

\section{Theoretical models} \label{sec:theory}

We consider a one-dimensional (1D) gas of $N$ bosons with repulsive contact interaction, which is  well-described by the quantum integrable Lieb-Liniger model~\cite{lieb1963exact, LL2}. The Hamiltonian in first-quantized form is given by
\begin{equation}
    \hat{\mathcal{H}} = - \sum _ { i } ^ { N } \frac { \hbar ^ { 2 } } { 2 m } \frac { \partial ^ { 2 } } { \partial z _ { i } ^ { 2 } } + g \sum _ { i < j } ^ { N } \delta \left( z _ { i } - z _ { j } \right) \; ,
    \label{eq:LiebLiniger_dimensional}
\end{equation}
where $m$ is the atomic mass and $g$ is the two-body coupling strength. The latter is, within the commonly used s-wave scattering approximation, directly related to the atomic scattering properties in cold atom experiments.

Several excellent reviews of Generalized Hydrodynamics (see Refs.~\cite{10.21468/SciPostPhysLectNotes.18, Bouchoule_2022, doyon2023generalized} for instance) and the Gross-Pitaevskii equation (see e.g.~Refs.~\cite{pitaevskii2003bose, pethick2008bose}) already exist.
Hence, we here restrict ourselves to only a short outline of the main concepts and results, relevant for the subsequent connection to the Whitham approach presented in Sec.~\ref{sec:Whitham}.
In particular, we will in the following focus on the zero-temperature case, i.e.~comparison between the mean-field GPE~\cite{pethick2008bose} and zero-entropy formulation of GHD~\cite{PhysRevLett.119.195301}.
We further note that we concern ourselves with the lowest order GHD (Euler-scale).

\subsection{Generalized Hydrodynamics}

Following the thermodynamics Bethe Ansatz, a local equilibrium macrostate of the system can be fully characterized by a distribution of quasi-particles $\rho_{\mathrm{p}}$~\cite{lieb1963exact, LL2}.
The quasi-particles are collective excitations of the system and exhibit fermionic statistics, with each particle being uniquely labelled by its quasi-momentum, or rapidity, $\theta$~\cite{PhysRevLett.19.1312}.
In the thermodynamic limit, the rapidity becomes a continuous variable.
Furthermore, if the system only varies on long space- and time-scales, one can assume that local equilibrium is always established.
Thus, the quasi-particle distribution can be considered time- and position-dependent $\rho_{\mathrm{p}} (z,t,\theta)$.
An equivalent representation of the macrostate is provided by the filling function $\vartheta(\theta) = \rho_{\mathrm{p}}(\theta) / \rho_{\mathrm{s}}(\theta)$, which describes the fraction of allowed rapidity states $\rho_{\mathrm{s}}(\theta)$ occupied by particles.

The ground state of the Lieb-Liniger model is given by a Fermi sea of rapidities, as the fermionic quasi-particles will fill up all low rapidity states up to some Fermi quasi-momentum $\Lambda_0$, thus yielding the filling function
\begin{equation}
    \vartheta(\theta) =
    \begin{cases}
        1, \qquad \text{for } -\Lambda_0 \leq \theta \leq \Lambda_0 \\
        0, \qquad \text{otherwise. }
    \end{cases}
\end{equation}
The value of $\Lambda_0$ is determined by the dimensionless parameter $\gamma = g m /(\hbar^2 n) = c/n$, where $n$ is the local atomic density and $c=g m/ \hbar^2$ is the re-scaled coupling strength.
In the presence of a finite hydrodynamic velocity $u$, the local Fermi quasi-momenta are boosted by $m u / \hbar$, yielding $\Lambda_\pm = m u / \hbar \pm \Lambda_0$.
Whilst the ground state and simple waves are represented by a single Fermi sea, more complicated states featuring multiple local Fermi seas are possible~\cite{PhysRevA.89.033637, Eliens_2016, 10.21468/SciPostPhys.1.1.008}.
For instance, the filling function of a state with two Fermi seas will assume the value 1 between the Fermi rapidity pairs $(\Lambda_1 , \Lambda_2)$ and $(\Lambda_3 , \Lambda_4)$, with 
$\Lambda_1 \leqslant \Lambda_2 \leqslant \Lambda_3 \leqslant \Lambda_4$, while 0 anywhere else. 
Thus, it is sufficient to encode any local zero-temperature state by only its Fermi rapidities.

Given the filling function $\vartheta(\theta)$, thermodynamic expectation values of local operators can be computed.
Of particular importance is the atomic density given by
\begin{equation}
    \bar n  = \frac{1}{2 \pi} \int_{-\infty}^{\infty} \mathrm{d} \theta \; \vartheta(\theta) \, (\partial_\theta p)_{\{ \Lambda \} }^{\mathrm{dr}}(\theta)  \; ,
    \label{eq:density_expectation_value_filling}
\end{equation}
and the hydrodynamic current
\begin{equation}
    \bar u  = \frac{1}{2 \pi \bar n} \int_{-\infty}^{\infty} \mathrm{d} \theta \; v_{\{ \Lambda \} }^{\mathrm{eff}}(\theta) \, \vartheta(\theta) \, (\partial_\theta p)_{\{ \Lambda \} }^{\mathrm{dr}}(\theta)  \; ,
    \label{eq:velocity_expectation_value_filling}
\end{equation}
where $p(\theta)$ is the momentum of a single quasi-particle with rapidity $\theta$.
Note that for a local state represented by multiple Fermi seas, the corresponding thermodynamic expectation values of operators, such as the ones above, remain single-valued and well-defined.
Further, the bar-notation indicates the coarse grained nature of the expectation values; in the presence of multiple Fermi seas, they correspond to the value obtained when averaging over the oscillation period of the emergent shock wave.

In equations \eqref{eq:density_expectation_value_filling} and \eqref{eq:velocity_expectation_value_filling} we have introduced the dressing operation defined as\begin{equation}
    f_{\{ \Lambda \} }^\mathrm{dr} (\theta) = f(\theta) + \frac{1}{2 \pi} \int_{-\infty}^{\infty} \mathrm{d}\theta'  \Delta(\theta,\theta') \vartheta (\theta') f_{\{ \Lambda \} }^\mathrm{dr} (\theta') \; ,
    \label{eq:dressing}
\end{equation}
where $\Delta(\theta,\theta') = \frac{2 c}{ c^2 + (\theta - \theta')^2}$ is the two-body scattering kernel of the Lieb-Liniger model.
The dressing equation reflects the inherent collective nature of integrable systems; interactions between all local quasi-particles lead to the modification of single-particle quantities.
Hence, for a zero-temperature state, dressed quantities are implicitly functions of the set of all local Fermi rapidities ${\{ \Lambda \} } = {\{ \Lambda_1, \Lambda_2,  \ldots \} }$, as denoted by the subscript.

\begin{figure}
    \centering
    \includegraphics[width = 1\columnwidth]{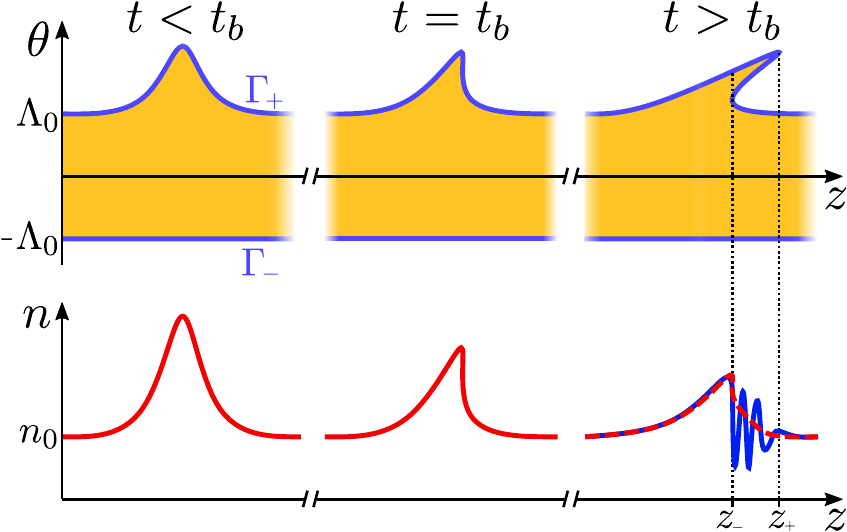}

    \caption{\label{fig:illustration}
    Illustration of the formation of a dispersive shock wave (DSW) resulting from a density perturbation propagating on a homogeneous condensate.
    On top, the Fermi contours $\Gamma_\pm$ of a zero temperature Lieb-Liniger state are shown, enclosing the rapidities $\theta$ for which the filling assumes $\vartheta = 1$, i.e.~the Fermi sea (shaded in orange).
    Outside the contour, $\vartheta = 0$.
    After the breaking time $t_b$, the local state in the region of the condensate where the DSW forms is represented by two Fermi seas.
    Below, the corresponding density is displayed; for $t < t_b $ the density varies slowly and is well-described by GHD (shown in red), while for $t = t_b$ the GPE solution develops a DSW (shown in blue).
    The length scale of the DSW oscillations are smaller than the coarse-grained scale of GHD, which therefore predicts the average of the DSW density. 
    }
\end{figure}

In an inhomogeneous system, the local Fermi rapidities of neighbouring states may vary. 
To parameterize the full system $\vartheta (z,t,\theta)$, we introduce the \textit{Fermi contours} $\Gamma_\pm (z,t)$, which are defined such that the filling is 1 for all points $(z,t)$ enclosed by the contours, and 0 outside.
For a single Fermi sea state, the contours are simply equal to the Fermi rapidities $\Lambda_\pm$, however, the contours can also be distorted into a locally multi-valued function to account for the presence of multiple Fermi seas. 
An example of this is illustrated in figure~\ref{fig:illustration}.
According to GHD, the dynamics of a zero-temperature state is given through the evolution of the contours following the dispersion-less equation~\cite{PhysRevLett.119.195301}
\begin{equation}
    \partial_t \Gamma_\pm + v_{\{ \Lambda \} }^\mathrm{eff} (\Gamma_\pm) \; \partial_z \Gamma_\pm = 0 \; .
    \label{eq:zero_temp_GHD}
\end{equation}
Here, the effective velocity is computed following~\cite{castro2016emergent, bertini2016transport, PhysRevX.10.011054}
\begin{equation}
    v_{\{ \Lambda \} }^\mathrm{eff} (\theta) = \frac{ (\partial_\theta \epsilon)_{\{ \Lambda \} }^\mathrm{dr} (\theta)}{ (\partial_\theta p)_{\{ \Lambda \} }^\mathrm{dr} (\theta)} \; ,   
    \label{eq:veff_contour}
\end{equation}
where $\epsilon(\theta) = \hbar^2 \theta^2 /(2m)$ and $p(\theta) = \hbar \theta$ are the single quasi-particle energy and momentum for the Lieb-Liniger model, respectively.
The effective velocity $v_{\{ \Lambda \} }^\mathrm{eff} (\theta)$ represents the ballistic propagation velocity of a quasi-particle with rapidity $\theta$; following interactions between particle their velocity is modified, which in the thermodynamic limit is captured by eq.~\eqref{eq:veff_contour}\cite{PhysRevX.10.011054}.
Meanwhile, in the semi-classical picture, the effective velocity can be understood as the accumulated effect of Wigner time delays associated with the phase shifts occurring under elastic collisions in quantum integrable systems~\cite{PhysRevB.97.045407,PhysRevLett.120.045301}.
Remarkably, it is possible to obtain eq.~\eqref{eq:veff_contour} from either perspective, thus demonstrating an exact quantum/classical correspondence~\cite{PhysRevX.10.011054, Borsi_2021}.

Notably, the evolution according to eq.~\eqref{eq:zero_temp_GHD} displaces the position of Fermi rapidities while preserving their value. 
Hence, to numerically simulate the GHD dynamics we discretize the contour as a set of points $\Gamma_\pm = \{ (z_j ,\Lambda_j) \}$, whose evolution is given by~\cite{PhysRevLett.119.195301}
\begin{equation}
    \frac{\mathrm{d} z_j }{\mathrm{d}t} = v_{\{ \Lambda \} }^\mathrm{eff} (\Lambda_j) \; ,
    \label{eq:zero_temp_GHD_discrete}
\end{equation}
where the effective velocity $v_{\{ \Lambda \} }^\mathrm{eff} (\Lambda_j) $ is evaluated by considering all local Fermi rapidities at position $z_j$.
Simulating the evolution of the contour is much easier than simulating the dynamics of the full filling function~\cite{10.21468/SciPostPhys.8.3.041, https://doi.org/10.48550/arxiv.2212.12349}.

\subsection{The Gross-Pitaevskii equation}

In the semi-classical limit, the Lieb-Liniger model is well approximated by the Gross-Pitaevskii equation (GPE)~\cite{novikov1984theory, pitaevskii2016bose}
\begin{equation}
    i \hbar \frac{\partial \psi}{\partial t}=-\frac{\hbar^2}{2 m} \partial_z^2 \psi + g|\psi|^2 \psi \; .
    \label{eq:GPE}
\end{equation}
The GPE describes the evolution of the order parameter $\psi(z,t)$ of a quasi-condensate. Rewriting the Lieb-Liniger Hamiltonian in second-quantized form, the GPE can be readily obtained as the equation of motion of the wave function for the condensate state, i.e.~replacing the creation- and annihilation operators by a classical field. 

By use of the Madelung representation of the order parameter \mbox{$\psi(z,t) = \sqrt{n(z,t)} e^{i \varphi (z,t)}$} in terms of the density $n(z,t)$ and phase $\varphi(z,t)$ variables, the GPE, Eq.~(\ref{eq:GPE}), can be formulated via an equivalent set of hydrodynamic equations
\begin{equation}
\begin{aligned}
& \partial_t n + \partial_z (n u)=0 \\
& \partial_t u +  \partial_z \left( \frac{1}{2} u^2 + \frac{g n}{m} - \frac{\hbar^2}{2 m^2} \frac{\partial_z^2 \sqrt{n}}{\sqrt{n}} \right)=0 \; ,
\end{aligned}
\label{eq:GPE_hydro_Madelung}
\end{equation}
where the hydrodynamic velocity field is defined as $u(z,t) = \frac{\hbar}{m}\partial_z \varphi(z,t)$. Eq.~(\ref{eq:GPE_hydro_Madelung}) is almost identical to the Euler equations, given the fact that the superfluid velocity is a potential flow, except for the last (dispersive) term $\sim \partial_z^2 \sqrt{n}$ called \emph{quantum pressure}. It is this term that is responsible for the absence of wave breaking in the GPE and the formation of the oscillatory wave train, since the quantum pressure dominates the usual pressure term for variations of the density on length scales less than the healing length $\xi_h = \hbar / \sqrt{2 m n g}$.

\section{The Whitham approach to GHD} \label{sec:Whitham}

An important feature of the effective velocity of the Lieb-Liniger model is that it is a monotonically increasing function of rapidity.
Hence, solutions with multiple local Fermi seas may develop dynamically from a single sea state~\cite{PhysRevLett.97.246402, Bettelheim_2008, PhysRevB.87.045112}; for instance following the "breaking" of a simple wave illustrated in figure~\ref{fig:illustration}, where the contour $\Gamma_+ (z,t)$ deforms over time to locally become multi-valued at the breaking time $t_b$.
The "breaking" of the Fermi contour heralds the development of a dispersive shock wave; in 1D a multi-valued density is entirely un-physical, whereby, as the density gradient approaches infinite, dispersion becomes necessary to regularize the solution.
The result is a non-linear, rapidly oscillating wave travelling in front of the breaking point $z_{-}$, namely a DSW.

The DSW theory of the GPE, also known as the Gurevich-Pitaevskii theory~\cite{1974JETP}, involves averaging over the periodic solution to obtain expressions for the parameters of a slowly modulated wave and its evolution. 
This slowly modulated wave, which connects the leading and trailing edges of the shock wave, can be represented by the so-called \textit{Riemann invariants}~\cite{10.1007/978-1-4613-8689-6_3, osti_5328912}.
Following Whitham theory, one can recover the microscopic structure of the DSW from the Riemann invariants~\cite{whitham2011linear}.
Meanwhile, GHD passes through the breaking point with seemingly no issue, although the theory fails to capture the oscillations of the DSW.
This is somewhat to be expected, as GHD is coarse-grained at a scale greater than the microscopic scale of the oscillations.
Indeed, GHD has been demonstrated to adequately capture the average density within the DSW \cite{PhysRevLett.125.180401, PhysRevResearch.3.013098}, as illustrated in figure~\ref{fig:illustration}.

For the 1D Bose gas, the Riemann invariants of the Whitham theory and the Fermi rapidities of GHD share many properties; for instance, their evolution is governed by the same type of dispersion-free equations.
Indeed, in the semi-classical limit, a correspondence between the theories was established in Ref.~\cite{Bettelheim_2020}.
In this section we expand upon the results of  Ref.~\cite{Bettelheim_2020}, explicitly stating the relation between Riemann invariants and Fermi rapidities and testing the agreement of the two theories approaching the semi-classical limit.
Then, in the following section, we combine Whitham's theory with GHD simulations to describe the formation of DSWs, thus expanding upon the capabilities of GHD.

\subsection{Equivalence in the dispersion-less limit}

The similarity between the Riemann invariants of Whitham theory and the Fermi contour of the zero-temperature GHD is apparent in the dispersion-less limit, where only large scale variations of the fluid are present.
In the absence of steep density gradients, the quantum pressure term of eq.~\eqref{eq:GPE_hydro_Madelung} can be neglected, whereby the hydrodynamic system assumes the form
\begin{equation}
\begin{aligned}
& \partial_t n + \partial_z (n u)=0 \; , \\
& \partial_t u +  u \partial_z u + \frac{g}{m} \partial_z n  =0 \; .
\end{aligned}
\end{equation}
Consider a simple wave propagating to the right, corresponding to the scenario in figure~\ref{fig:illustration} for $t < t_b$.
The flow velocity $u (z,t)$ and the density $n (z,t)$ can not change arbitrarily with respect to one another; instead, their relation is fixed by the Riemann invariants
\begin{equation}
    r_\pm = \frac{u}{2} \pm v_s \; ,
    \label{eq:dispersionless_Riemann_invariants}
\end{equation}
where $v_s = \frac{\hbar}{m} \sqrt{n c}$ is the sound velocity in a quasi-condensate.
Note we have omitted the $(z,t)$-dependence for a more compact notation.
Similarly to the Fermi contour of the zero-temperature GHD, the Riemann invariants are solutions that remain constant along the characteristics trajectories determined by the equations 
\begin{equation}
\partial_t r_{\pm} + v_{\pm}\partial_z r_{\pm} = 0 \; , 
\end{equation}
where the characteristic velocities $v_{\pm} = v_{\pm}\left(r_{+}, r_{-}\right)$ are given by
\begin{equation}
\begin{aligned}
    v_{+}=\frac{3}{2} r_{+}+\frac{1}{2} r_{-} \; ,\\
    v_{-}=\frac{1}{2} r_{+}+\frac{3}{2} r_{-} \; .
\end{aligned}
\label{eq:characteristic_velocities_displess}
\end{equation}
For a traveling wave, the Riemann invariants $r_\pm$ specify that $u$ and $v_s$ can not change arbitrary but must fulfil the relation~\eqref{eq:dispersionless_Riemann_invariants}.
Next, let us consider the Fermi contours $\Gamma_\pm (z,t)$ for the same scenario.
In the $\gamma \ll 1$ regime, the Fermi quasi-momentum of the ground state is accurately approximated by $\Lambda_0 = 2 n \sqrt{\gamma}$~\cite{lieb1963exact}.
Thus, it is straightforward to see that in the dispersion-less limit, the Fermi rapidities and the Riemann invariants are related by~\cite{PhysRevB.94.045110} 
\begin{equation}
    \Lambda_\pm = \frac{2 m}{\hbar} r_\pm \; .
    \label{eq:Fermi_Riemann_displess}
\end{equation}
Furthermore, the characteristic velocities of the GHD Fermi contour is given by the effective velocity $v^{\mathrm{eff}} (\Lambda_\pm)$.
Considering the microscopic definition of the sound velocity and the excitation spectrum of quasi-particles above the ground state~\cite{LL2}, one finds that the effective velocity at the Fermi edge is equal to the sound velocity~\cite{Cazalilla_2004}, that is $v_{\mathrm{s}} = v_0^{\mathrm{eff}}(\Lambda_0)$.
Thus, it is clear that $v^{\mathrm{eff}}(\Lambda_\pm) = v_\pm$ in the dispersion-less, semi-classical limit of the Lieb-Liniger model.

\subsection{Whitham modulation theory}

Similarly to the Fermi contour depicted in figure~\ref{fig:illustration}, following its breaking point a DSW is represented by four Riemann invariants $r_1 \leqslant r_2 \leqslant r_3 \leqslant r_4$ according to the Whitham modulation theory.
Details of the full procedure can be found in Refs.~\cite{kamchatnov2000nonlinear, whitham2011linear, EL201611}; here we report the Whitham equations, which describe the characteristic trajectories of the Riemann invariants, in their final form:
\begin{equation}
    \partial_t r_i + v_i \: \partial_z r_i = 0, \quad i=1,2,3,4 \; .
\end{equation}
The characteristic velocities $v_i = v_i(r_1, r_2, r_3, r_4)$ are given by the expressions
\begin{equation}    
\begin{aligned}
& v_1=V-\frac{\left(r_4-r_1\right)\left(r_2-r_1\right) K(M)}{\left(r_4-r_1\right) K(M)-\left(r_4-r_2\right) E(M)}, \\
& v_2=V+\frac{\left(r_3-r_2\right)\left(r_2-r_1\right) K(M)}{\left(r_3-r_2\right) K(M)-\left(r_3-r_1\right) E(M)}, \\
& v_3=V-\frac{\left(r_4-r_3\right)\left(r_3-r_2\right) K(M)}{\left(r_3-r_2\right) K(M)-\left(r_4-r_2\right) E(M)}, \\
& v_4=V+\frac{\left(r_4-r_3\right)\left(r_4-r_1\right) K(M)}{\left(r_4-r_1\right) K(M)-\left(r_3-r_1\right) E(M)} .
\end{aligned}
\label{eq:Whitham_system}
\end{equation}
where $V=\frac{1}{2}\left(r_1+r_2+r_3+r_4\right)$ is the velocity of the travelling wave, $K$ and $E$ are the complete elliptic integrals of first and second kind, respectively, and
\begin{equation}
    M=\frac{\left(r_2-r_1\right)\left(r_4-r_3\right)}{\left(r_4-r_2\right)\left(r_3-r_1\right)} \;, \quad 0 \leq M \leq 1 \; .
\end{equation}
Whitham modulation theory describes the DSW solution as a single-phased periodic wave with phase $\phi = k z - \omega t$.
The wave parameters (such as amplitude, wave-vector $k$, and frequency $\omega$) vary slowly and are parameterized by the Riemann invarants.
Thus, given the Riemann invariants, one can directly compute the periodic wave solution, whose density profile for the Gross-Pitaevskii equation reads
\begin{equation}
\begin{aligned}
n & =\frac{m^2}{4 c \hbar^2}\left(r_4-r_3-r_2+r_1\right)^2+ \frac{m^2}{c \hbar^2} \left(r_4-r_3\right) \\
& \times\left(r_2-r_1\right) \mathrm{sn}^2\left(  \phi, M\right)
\end{aligned}
\label{eq:density_Whitham}
\end{equation}
where $\phi = k z - \omega t + \phi_0$ is the DSW phase with the offset $\phi_0$, and $\mathrm{sn}(\cdot,M)$ is the Jacobi elliptic function.
The wave-vector $k$ and frequency $\omega$ are given by
\begin{equation}
    \begin{aligned}
    k &= \frac{m}{\hbar} \sqrt{\left(r_4-r_2\right)\left(r_3-r_1\right)} \; , \\
    \omega &= k V \; .
\end{aligned} 
\label{eq:DSW_wave_parameters}
\end{equation}
Similarly, the hydrodynamic velocity is given by
\begin{equation}
\begin{aligned}    
    u & = V - \frac{\hbar^2 c}{8 m^2 n} \left( -r_1 - r_2 + r_3 + r_4 \right) \\&\times \left( -r_1 + r_2 - r_3 + r_4 \right)  \left( r_1 - r_2 - r_3 + r_4 \right) \; .
\end{aligned}
\label{eq:velocity_Whitham}
\end{equation}
Note that in literature one typically defines convenient dimensionless variables, such that factors like $m^2 /(c \hbar^2 )$ in the expressions above vanish.\\

\begin{figure}
    \centering
    \includegraphics[width = 0.95\columnwidth]{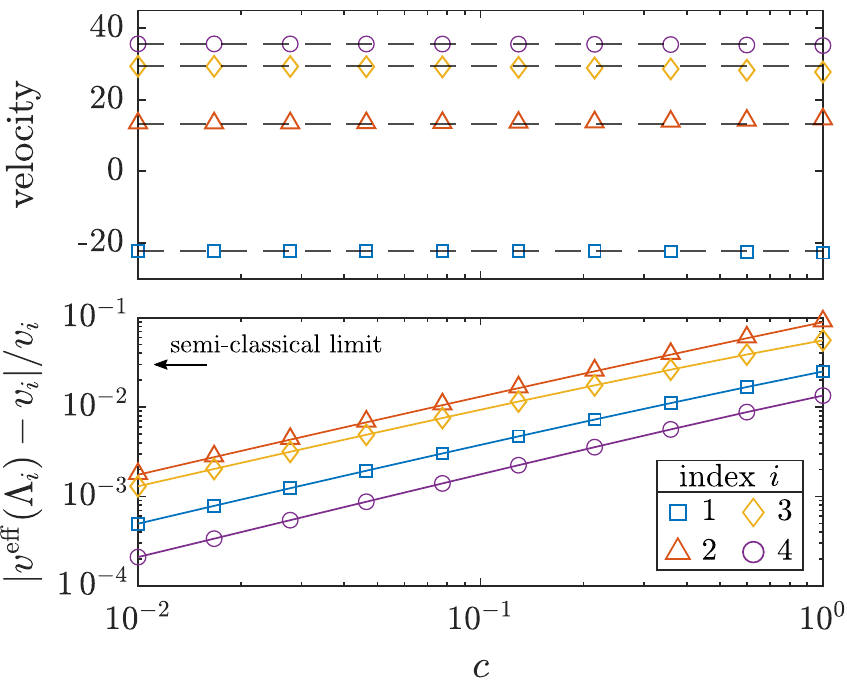}

    \caption{\label{fig:semiclassic_scaling}
    Top panel: The characteristic velocities $v_i$ of the Riemann invariants $r_1 = -20$, $r_2 = 15$, $r_3 = 20$, and $r_4 = 25$ are calculated via eqs.~\eqref{eq:Whitham_system} and plotted as black dashed lines.
    From the given Riemann invariants and varying coupling $c$, the Fermi rapidities $\Lambda_i$ are obtained via eq.~\eqref{eq:Fermi_Riemann} and their effective velocities are computed following eq.~\eqref{eq:veff_contour} and plotted as colored symbols.
    Bottom panel: Relative error between the two velocities as function of the coupling $c$.
    }
\end{figure}

Beyond the dispersion-less regime, establishing an explicit relation between the Riemann invariants and Fermi rapidities is a much more daunting task.
However, in the semi-classical limit of the one-dimensional Bose gas, the correspondence between Whitham's approach and GHD has been proven~\cite{Bettelheim_2020}.
This process involves finding to each quantum solution all conserved quantities and assigning to that solution a classical solution featuring the same set of conserved quantities.
Thus, an exact correspondence between the Riemann invariants and the Fermi rapidities of the classical and quantum solution, respectively, must exist, as it is indicative of a matching of the conserved quantities.
If governed by the same evolution equations, the Riemann invariants and Fermi rapidities must retain their relation in the dispersion-less regime~\eqref{eq:Fermi_Riemann_displess} past the formation of the DSW, whereby
\begin{equation}
    \Lambda_i = \frac{2 m}{\hbar} r_i \; , \quad \text{for } i = 1,2,3,4 \; .
    \label{eq:Fermi_Riemann}
\end{equation}
To test the convergence of the two theories, we directly compare the characteristic velocities~\eqref{eq:Whitham_system} with the effective velocity~\eqref{eq:veff_contour} while taking the semi-classical limit:
The semi-classical limit is accessed by introducing a fictitious Planck constant $h$ and scaling the quantum interaction and observables accordingly, such that $c = h c_{\mathrm{cl}}$ and $n = h^{-1} n_{\mathrm{cl}}$~\cite{De_Luca_2016, 10.21468/SciPostPhys.9.1.002} (see also Refs.~\cite{10.21468/SciPostPhys.4.6.045, PhysRevLett.125.240604, Koch_2022, bezzaz2023rapidity} specifically for semi-classical limits of GHD), with the subscript denoting the classical parameters; in the $h \to 0^+$ limit, i.e.~the semi-classical limit, the quantum system is well described by the classical model (GPE).
In practice, we perform the scaling by letting $c \to 0^+$ for a fixed set of Fermi rapidities (and thus Riemann invariants).
The resulting velocities are plotted in figure~\ref{fig:semiclassic_scaling}, expressed in units of $\hbar/ 2m = 1$.
For the given parameters, the product $c n \sim 80$ (where the density $n$ is computed using eq.~\eqref{eq:density_expectation_value_filling}) remains nearly constant, whereby the scaling conditions detailed above are met.
For $c \to 0^+$ we find that the effective velocities converge towards the characteristic velocities of the classical model; approaching the semi-classical limit, the errors scale as $\mathcal{O}(c)$.
The equivalence of the velocities in the semi-classical limit is to be expected given the results of Ref.~\cite{Bettelheim_2020}, however, we find that a reasonable agreement still persists in a regime accessible by current experimental setups~\cite{schemmer2019generalized, PhysRevX.12.041032}.
Therefore, assuming that we are in a regime with sufficient correspondence between the quantum and classical description, we can employ eq.~\eqref{eq:density_Whitham} to recover the periodic solution of DSWs directly from the Fermi contours of GHD.\\

Lastly, the Whitham approach has also been employed to study density oscillations following a shock in the 1D free fermion gas~\cite{PhysRevB.87.045112}, whose dynamics is equivalent to the 1D Bose gas in the strongly interacting Tonks-Girardeau regime ($\gamma \gg 1$, i.e.~ opposite to the regime treated here).
Notably, the connection between GHD and Whitham's theory is much clearer in the Tonks limit, as the Wigner function of free fermion gas and its dynamics are identical to the filling function $\vartheta$ and eq.~\eqref{eq:zero_temp_GHD}.
However, in strongly interacting systems, the density oscillations look very different to those of the GPE: Where the latter features large amplitudes with little resemblance of the mean density, quantum simulations of free fermions produce much smaller oscillations around the mean (see Refs.~\cite{PhysRevLett.125.180401, https://doi.org/10.48550/arxiv.2208.06614, PhysRevB.87.045112} for examples).
Hence, the Whitham equations of free fermions look rather different from those of the GPE.
In Appendix~\ref{app:TG_regime} we present the hydrodynamics of the former as derived in Ref.~\cite{PhysRevB.87.045112} and directly compare the Whitham approaches for the two opposite regimes (see also the discussion in Sec.~\ref{sec:discussion}).
Summarizing the results, the oscillation period of the free fermion hydrodynamics is determined by the wave vector $k_{3 2} = \Lambda_3 - \Lambda_2$, and is thus parametrically different from that predicted by the GPE via eq.~\eqref{eq:density_Whitham}.
Notably, the oscillation period for free fermions is completely independent of the lower Fermi contour.
This is reflective of how many-body interactions manifest differently in the strongly ($\gamma \gg 1$) and weakly interacting ($\gamma \ll 1$) limits of the 1D Bose gas.
One can see this clearly when considering the dressing equation~\eqref{eq:dressing}; for $c \to \infty$ the two-body scattering kernel becomes vanishing and $ f^\mathrm{dr} (\theta) \to f(\theta)$.
Thus, for in the strongly interacting regime, the presence of an additional local Fermi sea barely changes the Bethe Ansatz quasi-particle distribution $\rho_{\mathrm{p}} (\theta)$ of the background state (and vice versa).
Meanwhile, for $\gamma \ll 1$, the quasi-particle distributions corresponding to separate Fermi seas are modified from their single Fermi sea counterpart.
Finally, it should be noted that the free fermion hydrodynamics of Ref.~\cite{PhysRevB.87.045112} is derived in a phenomenological manner; although it provides an improved description when compared to the GPE, it still fails to capture features seen in results from direct quantum simulations.

\section{The bipartition problem} \label{sec:Riemann_problem}

\begin{figure*}[t]
    \centering
    \includegraphics[width = 1\textwidth]{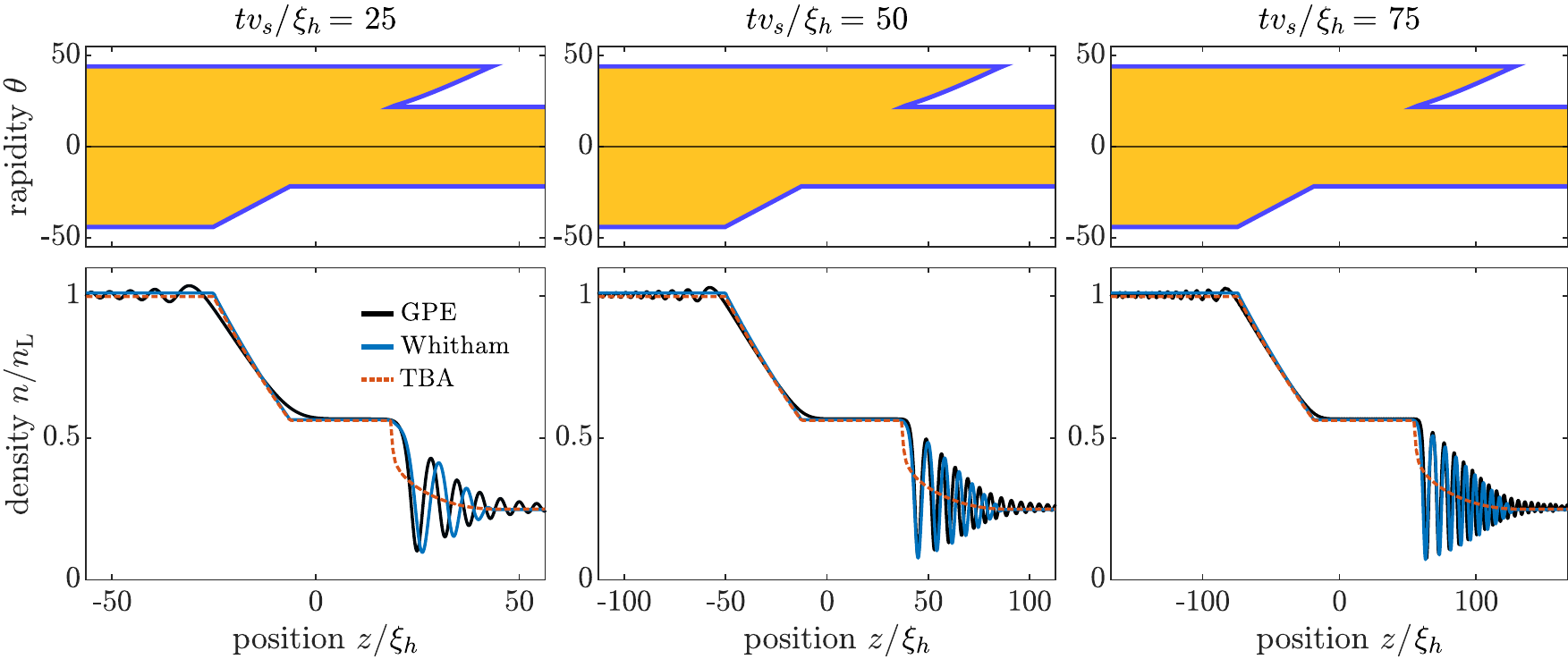}

    \caption{\label{fig:Riemann_snapshots}
    Wave structure formed following evolution of an initial density discontinuity.
    Top row: Zero temperature Lieb-Liniger state represented by the Fermi contours $\Gamma_\pm$, whose evolution is simulated using GHD.
    Bottom row: Given the simulated Fermi contours, Whitham's approach is used to calculate the atomic density via eq.~\eqref{eq:density_Whitham}. 
    For comparison, the results of numerical GPE simulations are plotted.
    Finally, the coarse grained expectation value of the density given by thermodynamics Bethe Ansatz (TBA) $\bar n$ of eq.~\eqref{eq:density_expectation_value_filling} is plotted, capturing the average density of the DSW.
    }
\end{figure*}

One of the simplest setups for studying dispersive shock waves is the bipartition problem, where an infinitely long system features an initial discontinuity at $z=0$.
Here we consider an initial discontinuity in the density profile with zero initial hydrodynamic velocity, specifically $u_\mathrm{L} = u_\mathrm{R} = 0$ and $n_\mathrm{L} > n_\mathrm{R}$.
At $t>0$ the solution of the GPE for the given initial condition consists of two waves traveling in opposite directions; on the side with higher density a rarefaction wave forms, while on the low-density side a DSW forms, here travelling to the right~\cite{EL1995186}.
The two waves propagate away from the point of the initial discontinuity and are joined by a domain of homogeneous flow, where a density plateau forms.

We consider a setup with initial densities $n_\mathrm{L} = 1000$ and $n_\mathrm{R} = 250$ and a coupling strength $c = 0.5$; this is sufficiently close to the semi-classical limit to achieve a good agreement between the quantum and classical model.
The Lieb-Liniger parameters of the left and right side are $\gamma_L = 0.0005$ and $\gamma_R = 0.002$, respectively, whereby the initial Fermi quasi-momenta are accurately given by $\Lambda_0 = 2 n \sqrt{\gamma}$.
It is convenient to define appropriate length and time scales to express the results; here we employ the healing length $\xi_h =  1/\sqrt{2 n_L c}$ and sound velocity $v_s = \hbar \sqrt{n_L c} /m$ evaluated according to the density on the left boundary.

\subsection{Zero temperature} \label{sec:zero_temp}

At zero temperature, the GHD dynamics is solved numerically by propagating the Fermi contours $\Gamma_\pm$ according to eq.~\eqref{eq:zero_temp_GHD_discrete}.
In figure~\ref{fig:Riemann_snapshots} we plot Fermi contours (and associated filling function) at select evolution times.
On the right side of the initial discontinuity ($z > 0$), a region has developed in which the state is represented by two Fermi seas, thus indicating the formation of a DSW.
Meanwhile, the wave traveling to the left, represented by a single, widening Fermi sea, is a rarefaction wave.
Comparing with typical illustrations of Riemann invariants for this problem (see Ref.~\cite{Kamchatnov_2021} for instance), we find a high resemblance to the Fermi contours of our simulation.

Next, given the propagated Fermi contours, we calculate the corresponding Riemann invariants via eq.~\eqref{eq:Fermi_Riemann} from which we compute the density of the traveling wave solution according to Whitham theory~\eqref{eq:density_Whitham}.
The results are plotted in figure~\ref{fig:Riemann_snapshots}.
In the region of the system where the local state is described by two Fermi seas, the calculated density exhibits the rapidly oscillating behaviour of a DSW.
We stress again that this is beyond the capabilities of the thermodynamic Bethe Ansatz, whose coarse grained expectation value for the density~\eqref{eq:density_expectation_value_filling} is plotted for comparison, which captures only the average of the oscillating solution.

Lastly, we compare our results with numerical simulations of the GPE system~\eqref{eq:GPE_hydro_Madelung} employing the Fourier split-step method; the results are plotted in figure~\ref{fig:Riemann_snapshots}.
We find a good agreement upon comparing the two densities, particularly when considering that the GHD simulations are conducted at finite $c$.
Notably, the agreement within the DSW region becomes better at longer evolution times. 
The reason therefore is twofold:
Firstly, the Riemann approach is only expected to be accurate in the asymptotic limit, where a clear separation of scales between the oscillations of the DSW and their modulation is found.
Secondly, the GPE simulations are initialized with a smooth, but steep boundary between the two halves of the system such that typical length scales of the problem can be faithfully represented on the numerical grid with finite resolution.
Hence, at shorter times, artifacts of the boundary may still be present in the solution.

In the appendix, we present further demonstrations of the application of Whitham's theory to GHD simulations:
Appendix~\ref{app:Riemann_background_variantion} demonstrates the dependence of the DSW shape on the background density, while Appendix~\ref{sec:piston_problem} features benchmarks on the problem of a moving piston.
In both cases, the results of calculating the DSW density~\eqref{eq:density_Whitham} using the simulated GHD contour exhibit good agreement with GPE simulations.

\subsection{Finite temperature}

Finite temperature effects are known to wash out the contrast of DSWs~\cite{PhysRevLett.125.180401}.
In Whitham modulation theory, we find that thermal effects manifest as incoherent fluctuations of the DSW phase $\phi$, resulting in a dephasing of the oscillating solution.
To see this, consider the collective variables of density and hydrodynamic velocity, which are subject to thermal fluctuations following the bosonization procedure~\cite{Haldane81, Haldane94, Cazalilla_2004}.
These fluctuations are superpositions of many thermally populated momentum modes; for thermal states with low temperature $T$, we can assume that the thermal fluctuations are statistically mutually independent, whereby their probability follow a normal distribution with zero mean and variance
\begin{align}
    \sigma_{\delta n}^2 &=\frac{\mathcal{K}}{6}\left(\frac{k_{\mathrm{B}} T}{\hbar v_{\mathrm{s}} }\right)^2  \; ,\\
    \sigma_{\delta u}^2 &= \frac{\hbar^2 \pi^2}{6 m^2 \mathcal{K}}\left(\frac{k_{\mathrm{B}} T}{\hbar v_{\mathrm{s}} }\right)^2 \; ,
\end{align}
where $\mathcal{K} = 2 \pi n/v_s$ is the Luttinger liquid parameter.
The fluctuations of density and velocity result in fluctuations of the Fermi contours, or equivalently, the Riemann invariants $\delta r_\pm$; following the central limit theorem, fluctuations of the Riemann invariants are also Gaussian with zero mean, and their variance can be obtained using equations \eqref{eq:dispersionless_Riemann_invariants} and \eqref{eq:Fermi_Riemann_displess} yielding
\begin{equation}
    \sigma_{\delta r_\pm}^2 = \frac{1}{4} \sigma_{\delta u}^2 + \frac{\hbar^2 \gamma}{4 m^2} \sigma_{\delta n}^2 \; .
\end{equation}
Next, to understand how fluctuations affect the contrast of DSWs, we assume that the fluctuations are sufficiently small to neglect their contribution to the effective velocity, i.e.~the dynamics of fluctuations is linearized around the background state similarly to time-dependent Bogoliubov theory or Quantum GHD.
Thus, the distribution of fluctuations for each Riemann invariant is preserved following the formation of a shock, and the variance of DSW phase fluctuations $\delta\phi$ reads
\begin{equation}
    \sigma_{\delta \phi}^2 = \sum_{i = 1}^{4} \left( \frac{\partial \phi}{\partial r_i} \right)^2 \sigma_{\delta r_i}^2 \; .
\end{equation}
From this expression, we see that the variance of the DSW phase following the shock formation scales with $T^2$ of the initial state.
Meanwhile, the partial derivatives depend on details of the oscillating solution and can be worked out from eq.~\eqref{eq:DSW_wave_parameters} yielding
\begin{equation}
    \frac{\partial \phi}{\partial r_i} = \frac{\partial k}{\partial r_i} \left( z - V t \right) - \frac{k}{2}t \; ,
    \label{eq:DSWphase_derivative}
\end{equation}
with derivatives of the wave-vector $k$ being
\begin{equation}
\begin{aligned}
    \frac{\partial k}{\partial r_1} &= \frac{- (r_4 -r_2) m^2}{2 \hbar^2 k} \; , \\
    \frac{\partial k}{\partial r_2} &= \frac{- (r_3 -r_1) m^2}{2 \hbar^2 k} \; , \\
    \frac{\partial k}{\partial r_3} &= \frac{ (r_4 -r_2) m^2}{2 \hbar^2 k} \; , \\
    \frac{\partial k}{\partial r_4} &= \frac{ (r_3 -r_1) m^2}{2 \hbar^2 k} \; .
\end{aligned}
\label{eq:wavevector_derivative}
\end{equation}
Consider the bipartition problem of the previous section, where the two halves of the system are initially in thermal equilibrium with one another, i.e.~they share the same temperature $T$.
Following the shock ($t >0$), the Riemann invariant fluctuations $\delta r_1$ and $\delta r_2$ are determined by the initial density on the right boundary $n_\mathrm{R}$, while $\delta r_4$ depends on $n_\mathrm{L}$.
The Riemann invariant $r_3$ corresponds to the initial boundary region between the two halves of the system; as this region is infinitesimally small, the thermal fluctuations, which at low temperatures are dominated by long wavelength modes, are vanishing.
Hence, equations \eqref{eq:DSWphase_derivative} and \eqref{eq:wavevector_derivative} show that $\sigma_{\delta \phi}^2$ is greater for larger density difference between the two system halves.
Furthermore, the contribution of thermal fluctuations is strongest around the front edge of the DSW where the difference $r_3 - r_1$ is maximal (see figure~\ref{fig:Riemann_snapshots} for reference).

\begin{figure}
    \centering
    \includegraphics[width = 0.9\columnwidth]{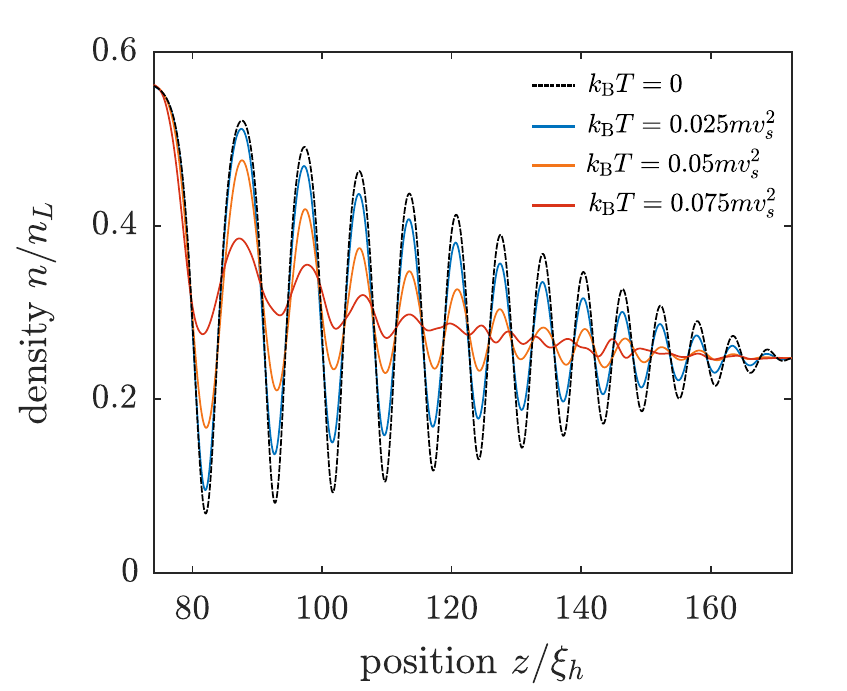}

    \caption{\label{fig:Riemann_finite_temperature}
    Reduction of DSW contrast following thermal fluctuations in the bipartition problem. Finite temperature states are sampled following Ref.~\cite{Moller2022}, then propagated according to GHD~\eqref{eq:zero_temp_GHD}. The evolution time is $t v_s /\xi_h = 75$. The density is calculated using Whitham's approach~\eqref{eq:density_Whitham} and averaged over the ensemble.
    }
\end{figure}

To numerically test these predictions, we proceed in the spirit of the truncated Wigner approximation, following the approach of Ref.~\cite{Moller2022}:
Starting from a given zero temperature state $\Gamma_{\pm}^{T = 0}$ (here considering the bipartition problem of the previous section), we sample density and phase fluctuations for a given temperature $T$ and then re-solve the Bethe Ansatz to obtain the fluctuating Fermi contours $\Gamma_{\pm}^{T}  (z) = \Gamma_{\pm}^{T = 0} + \delta\Gamma_\pm (z)$.
Many initial states of fluctuating Fermi contours are independently sampled (here 200), such that the population of each collective mode follows the Bose-Einstein distribution (fluctuation quasi-particles "boostons" are bosonic).
Next, each realization is propagated according to a linearized GHD, following eq.~\eqref{eq:zero_temp_GHD} where the contribution of fluctuations to the effective velocity is neglected.
Lastly, observables (here the DSW density via eq.~\eqref{eq:density_Whitham}) are evaluated for each realization, then averaged over the ensemble to produce the final result.
Results obtained for different temperatures are shown in figure~\ref{fig:Riemann_finite_temperature}: As predicted, the DSW contrast is reduced as temperature increases, and we find a stronger distortion of the DSW towards its front end, particularly for the hotter realizations.

For comparison, simulations of the finite temperature bipartition using the stochastic-projected GPE method were carried out; initial states with thermal fluctuations are acquired using a stochastic noise term, then propagated according to zero-temperature GPE~\eqref{eq:GPE_hydro_Madelung}. 
Here we find the effect of thermal fluctuations weaker than in GHD, i.e.~a higher temperature is required to achieve the same level of contrast reduction as seen in figure~\ref{fig:Riemann_finite_temperature}.
We attribute this discrepancy to the GPE dispersive term coupling to the fluctuations:
When not linearizing the fluctuation dynamics, the fluctuations themselves are known to create tiny shocks~\cite{Moller2022}; dispersively regularizing these shocks introduces oscillations with frequency beyond the UV cutoff of the GPE simulation, effectively diminishing the fluctuations.
Notably, this mechanism is especially prominent within the region of the main shock.
A detailed analysis of linearized and non-linear fluctuation dynamics is an interesting perspective for further studies, but beyond the scope of this work.

\subsection{Shock wave evolution during free expansion}

In experiments with ultracold atoms, a routine method of probing the system is releasing the trapping potential, thus effectively quenching interactions to zero, and letting the atomic cloud undergo free expansion followed by imaging of its density.
In one-dimensional systems, such expansion protocols lead to the formation of \textit{density ripples} along the longitudinal axis, as fluctuations of the condensate phase (hydrodynamic velocity field) transform into density fluctuations~\cite{PhysRevA.80.033604}.
These phase fluctuations are typically thermal, leading to density ripples being used for thermometry by measuring their (temperature dependent) power spectrum~\cite{PhysRevA.81.031610} or by training neural networks to associate ripple features to the underlying temperature~\cite{PhysRevA.104.043305}.
Additionally, density ripples have been employed to monitor squeezed collective modes~\cite{PhysRevA.98.043604}.

\begin{figure}
    \centering
    \includegraphics[width = 1\columnwidth]{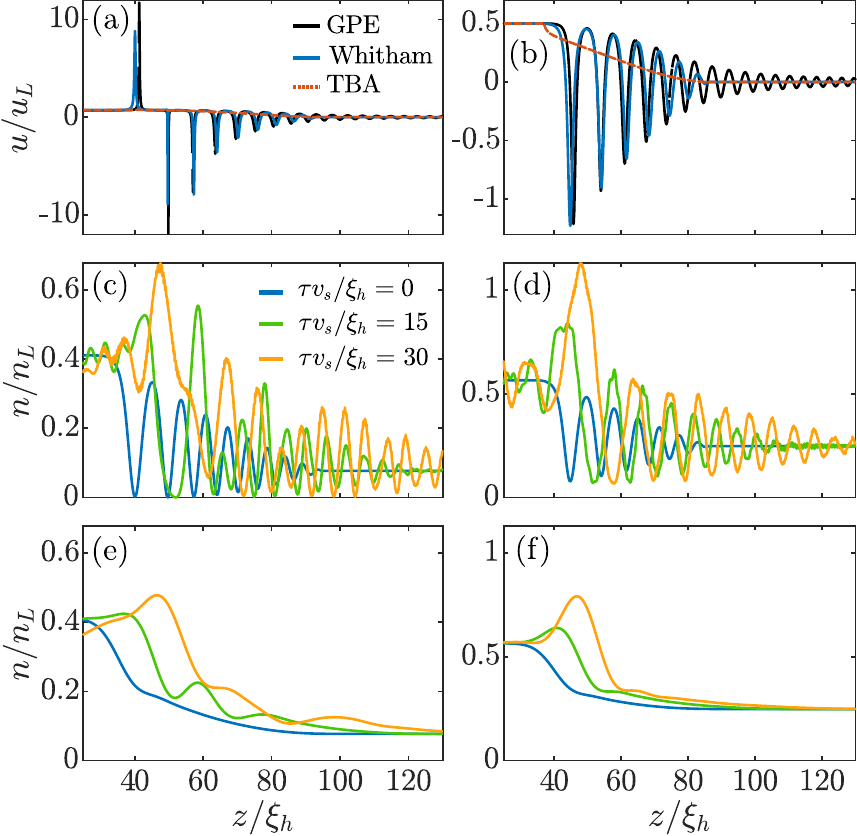}
    \phantomsubfloat{\label{fig:DSW_TOF_a}}
    \phantomsubfloat{\label{fig:DSW_TOF_b}}
    \phantomsubfloat{\label{fig:DSW_TOF_c}}
    \phantomsubfloat{\label{fig:DSW_TOF_d}}
    \phantomsubfloat{\label{fig:DSW_TOF_e}}
    \phantomsubfloat{\label{fig:DSW_TOF_f}}
    \vspace{-2\baselineskip}

    \caption{\label{fig:DSW_TOF}
    Evolution of Riemann protocol DSW during free expansion.
    \textbf{(a, b)} Hydrodynamic velocity obtained from zero-temperature GHD simulations using TBA~\eqref{eq:velocity_expectation_value_filling} and Whitham's approach~\eqref{eq:velocity_Whitham} for right-hand density of $n_\mathrm{R} = 0.08 n_\mathrm{L}$ and $n_\mathrm{R} = 0.25 n_\mathrm{L}$, respectively. The simulation duration is $t v_s /\xi_h = 50$. The results are compared to GPE simulations. 
    \textbf{(c, d)} Corresponding density profiles following free evolution of duration $\tau$.
    \textbf{(e, f)} Convolution of the density profiles with a Gaussian with standard deviation $\sigma = 5 \xi_h$.
    }
\end{figure}

Following the formation of a DSW, large oscillations in the hydrodynamic velocity (in addition to the oscillating density) develop as per eq.~\eqref{eq:velocity_Whitham}.
Thus, during free expansion of an atomic cloud, features of the shock wave are magnified, which could be highly beneficial for the purpose of experimentally detecting them.
Indeed, in experiments realizing a Lieb-Liniger gas in the weakly interacting regime ($\gamma \ll 1$), the imaging resolution is typically much larger than the wavelength of the shock wave.
Hence, density profiles measured in-\textit{situ} resemble the TBA prediction~\eqref{eq:density_expectation_value_filling}, as the imaging apparatus effectively averages over the oscillating shock wave density.

In figure~\ref{fig:DSW_TOF} we illustrate the amplification of shock wave features following free expansion.
Figures~\ref{fig:DSW_TOF_a} and \ref{fig:DSW_TOF_b} show the hydrodynamic velocity for the zero-temperature bipartition problem; the latter subfigure corresponds to the system also depicted in figure~\ref{fig:Riemann_snapshots}, while the former is calculated for a smaller right-hand density $n_\mathrm{R} = 0.08 n_\mathrm{L}$.
Starting from the Fermi rapidities of the GHD simulation, the velocity is computed using TBA~\eqref{eq:velocity_expectation_value_filling} and Whitham's approach~\eqref{eq:velocity_Whitham}. 
When the shock wave amplitude approaches the background density, the hydrodynamic velocity features very sharp peaks at the locations of the density minima.
Again, we find a good agreement with GPE simulations.

Next, in figures~\ref{fig:DSW_TOF_c} and \ref{fig:DSW_TOF_d}, the shock wave density following different free expansion times $\tau$ is depicted for the two Riemann setups above.
For $\tau > 0$, the shock wave continues propagating towards the right while experiencing a substantial growth in its contrast, particularly around the back of the wave where the magnitude of the velocity is the largest.
Interestingly, although the hydrodynamic velocities of the two setups are very different, their density profiles following free expansion are rather similar.

Finally, to emulate the effect of an imaging system, we convolve the freely evolved density profiles with a Gaussian.
Here we have selected a standard deviation $\sigma = 5 \xi_h$, which is representative of the resolution found in experimental setups.
The results are plotted in figures~\ref{fig:DSW_TOF_e} and \ref{fig:DSW_TOF_f}.
For $\tau = 0$, corresponding to an in-\textit{situ} measurement, we indeed obtain a density profile very similar to the TBA prediction. 
However, for $\tau > 0$, the magnified features of the shock wave results in a noticeable density peak near the back of the shock wave, which continues to grow for increasing expansion times.
In the $n_\mathrm{R} = 0.08 n_\mathrm{L}$ case, even secondary peaks towards the center of the shock wave appear in the "imaged" profiles.
While thermal effects still represent a significant challenge in the context of measuring DSWs, the results of figure~\ref{fig:DSW_TOF} are encouraging.

\section{Towards generalization to arbitrary interaction strength} \label{sec:discussion}

The results presented here lend themselves to a clear physical interpretation of the formation of DSWs in the 1D Bose gas.
For a state represented by a single Fermi sea and where $\gamma \ll 1$, the local density is determined by the squared width of the Fermi sea, i.e.~ the squared difference of its Fermi rapidities.
Meanwhile, in the presence of multiple local Fermi seas, their contributions \textit{interfere} rather than simply add up, which can easily be seen from eq.~\eqref{eq:density_Whitham}; its first term represents the lower envelope of the shock wave density, i.e.~ the destructive interference of the two components.
The interpretation of DSWs as the interference of two local fluid components represented by separate Fermi seas complements the conclusion of Ref.~\cite{PhysRevLett.125.180401}, namely that DSWs are the result of quantum mechanical self-interference between a traveling wave and its background.
A similar conclusion was drawn in the study of shock wave formation in free Fermi gases of Ref.~\cite{PhysRevB.87.045112}, where the oscillations were interpreted as Friedel-type oscillations between different branches of the Fermi momentum.

However, the picture facilitated by Whitham's theory combined with the Bethe Ansatz extends past that of a perturbation propagating on a background; indeed, the GHD description goes beyond mean field dynamics, extending to far out-of-equilibrium setups like the quantum Newton's cradle~\cite{kinoshita2006quantum, 10.21468/SciPostPhys.6.6.070, 10.21468/SciPostPhys.9.4.058}.
Furthermore, since GHD of the Lieb-Liniger model is built upon its exact Bethe Ansatz solutions, the theory applies for all interaction strengths $\gamma$ (and temperatures).
This raises the question, whether the ideas of Whitham's approach can be framed in a more general scenario, thus facilitating a description in any regime with the Fermi rapidities constituting a set of \textit{generalized} Riemann invariants.
Indeed, the effective velocity~\eqref{eq:veff_contour} correctly describes the characteristic velocities of such generalized Riemann invariants for all interaction strengths, however, no general expression for the periodic solution of the DSW exists as of yet.
The shape of the shock wave depends on the underlying dispersive hydrodynamics; as discussed in Section~\ref{sec:Whitham} (see also Appendix~\ref{app:TG_regime}), the Whitham GPE does not extend to strongly interacting regime.
To obtain the correct dispersive hydrodynamics (at arbitrary interaction strength), one would need to derive the dispersive term directly from the Bethe Ansatz, possibly following the approach outlined in Ref.~\cite{denardis2022hydrodynamic}, and then apply Whitham’s theory.

Further, the combined picture has clear analogies to the Quantum Generalized Hydrodynamics (QGHD) of Ref.~\cite{PhysRevLett.124.140603}; in fact, there are indications that a Whitham's theory of GHD could be seen as the semi-classical limit of QGHD. 
Conceptually, QGHD is a generalization of time-dependent and inhomogeneous Luttinger Liquids to truly out-of-equilibrium situations.
Luttinger Liquids essentially describe (quantum) fluctuations of the (single) Fermi sea~\cite{giamarchi2003quantum, Cazalilla_2004}, whilst QGHD captures the action for the fluctuations when placed on top of the evolving GHD.
The formalism enables the connection with physical observables such as correlations through operator expansion~\cite{10.21468/SciPostPhys.2.2.012, 10.21468/SciPostPhys.4.6.037}, although these calculations remain notoriously difficult for systems with large numbers of atoms.
However, in the strongly interacting Tonks-Girardeau regime, the derivation of asymptotically exact formula is possible, which has enabled QGHD to describe Friedel oscillations of the density profile in setups with multiple local Fermi seas~\cite{Ruggiero_2022, PhysRevA.108.013324}.
In fact, QGHD correctly captures the dynamics of direct quantum simulations in the strongly interacting regime~\cite{private_comm_stefano}, like the ones of Ref.~\cite{PhysRevB.87.045112}. 
By explicitly connecting QGHD and Whitham's theory, the latter could provide a prescription to fix non-universal prefactors in the operator expansions (at least close to the semi-classical limit), thereby aiding in future developments of QGHD and providing important benchmarks. 
Establishing such a connection is the beyond the scope of this work, however, we will outline some of the considerations needed to be made in this regard.
The first step would be to match the oscillation wave vectors. 
Within the Whitham approach, one considers the solution to hydrodynamic equations in the shock region as a single-phase periodic wave with slowly modulated parameter (such as the wave vector).
To this end, the single wave vector is expressed in terms of the Riemann invariants (here Fermi rapidities), whose evolution is governed by a set of dispersion-less equations.
Direct quantum simulations in the strongly interacting regime reveal a much more irregular oscillating pattern; the solution is thus not single-phased as it instead features contributions from multiple interferences between different Fermi points~\cite{PhysRevB.87.045112, PhysRevA.108.013324}.
Some of these contribution may be strictly quantum mechanical and thus not captured by hydrodynamics; in order to attempt to match the different approaches, one would first need to understand the full extend of such discrepancies.

\section{Conclusion} \label{sec:conclusion}

In this work, we have studied the connection between Whitham modulation theory and Generalized Hydrodynamics.
To this end, we have first explicitly related the Riemann invariants of Whitham's theory for the GPE to the Fermi rapidities of the Lieb-Liniger model and tested the convergence of the theories approaching the semi-classical limit.
Here we find a good agreement in regimes accessible in current experiments with ultracold quantum gases.
Next, we have demonstrated how short wavelength density oscillations following a hydrodynamic shock, otherwise lost in the GHD description, can be recovered through Whitham's approach.
We have further shown how thermal fluctuations diminish the contrast of the shock wave, whilst free expansion of the system amplifies its features and thus facilitates experimental detection.
At zero temperature, our results are in agreement with GPE simulations; at finite temperature, the qualitative behavior is the same, although we find a higher temperature necessary for GPE to achieve the same level of contrast reduction.

The combined picture of the two theories lends itself to additional physical interpretation, connecting the formation of DSWs to the interference between multiple Fermi seas representing separate components of the fluid.
Additionally, it exhibits clear parallels to the theory of Quantum Generalized Hydrodynamics; explicitly connecting Whitham's theory to QGHD could provide a prescription to fix non-universal prefactors in operator expansion of the latter.
However, establishing said connection at arbitrary interaction strength entails several challenges: 
In the strongly interacting regime $\gamma \gg 1$, where successful operator expansions have been made, the correct dispersive hydrodynamic theory is yet to be determined. 
To achieve this (at arbitrary interaction strength), one would likely need to derive the dispersive hydrodynamics directly from the Bethe Ansatz.
Alternatively, in the opposite regime, operator expansions of the density operator could be aided by comparisons with the known Whitham modulation equations of the GPE; here, Whitham's theory could help identify the most relevant terms of the expansion.

\section*{Acknowledgements}

We thank Igor Mazets, Eldad Bettelheim, J\'er\^ome Dubail and Stefano Scopa for useful discussions. 
This work has been supported by the DFG/FWF CRC 1225 'Isoquant',  the DFG/FWF Research Unit FOR 2724 'Thermal machines in the quantum world', and the FQXi program on 'Information as fuel' ESQ Discovery Grant 'Emergence of physical laws: from mathematical foundations to applications in many-body physics' of the Austrian Academy of Sciences (O{\"A}W).

\appendix

\section{Whitham's approach to the 1D free fermion gas} \label{app:TG_regime}

\begin{figure}
    \centering
    \includegraphics[width = 1\columnwidth]{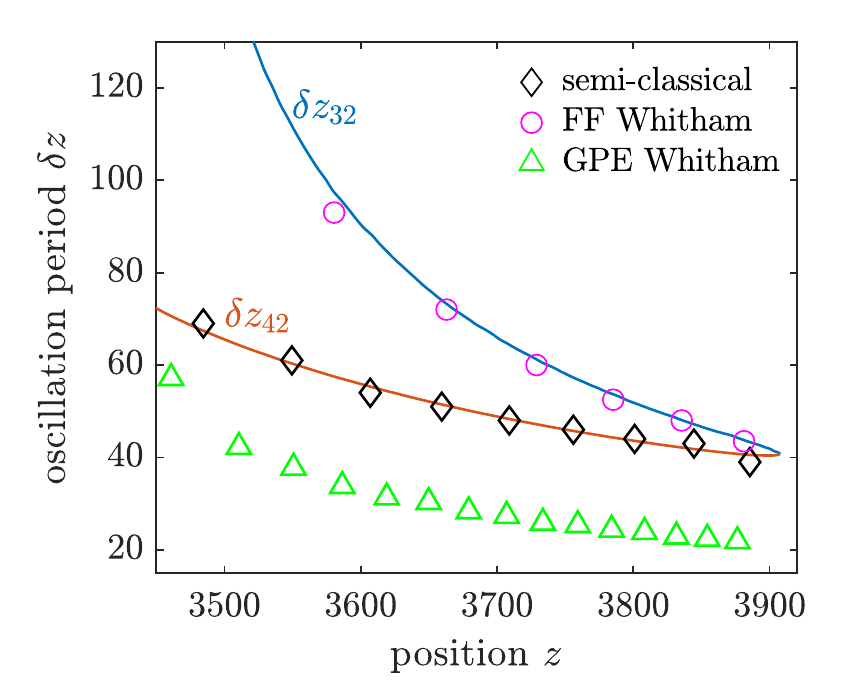}

    \caption{\label{fig:TG_Whitham}
    Oscillation period of a shock wave from a density perturbation in the 1D free Fermi gas. 
    Data is taken from Ref.~\cite{PhysRevB.87.045112}.
    Symbols denote the periods obtained via: Evolution of the semi-classical Wigner function (diamonds), Whitham theory of an approximate hydrodynamics of the free Fermi gas~\eqref{eq:densit_Whitham_FF} (circles), and the Whitham theory for the Gross-Pitaevskii equation~\eqref{eq:density_Whitham} (triangles).
    The lines indicate the period associated with the wave vector of a particular momentum branch interference, i.e.~ $\delta z_{ij} = 2 \pi /(\Lambda_i - \Lambda_j)$.
    }
\end{figure}

In this section, we present the modulation equations for an approximate hydrodynamic theory for the free Fermi gas, originally derived in Ref.~\cite{PhysRevB.87.045112} (see also Refs.~\cite{PhysRevLett.74.5153, Gutman2007}) in the context of a density perturbation travelling on a homogeneous background.

The starting point for obtaining the hydrodynamics is the Euler equation for the free Fermi (FF) gas Wigner function, which notably is identical to the zero temperature GHD of the Tonks-Girardeau gas.
Next, the equation is rephrased in terms of the mean density and velocity of the fluid and a dispersive term is included to regularize the shock.
To this end, two regularizing contributions are considered, namely gradient corrections to the Hamiltonian and loop corrections.
The resulting hydrodynamic equations read 
\begin{equation}
\begin{aligned}
    &\partial_t n + \partial_z( n u )=0 \; \\
    &\partial_t u + u \partial_z u + \partial_z w = 0 \; ,    
\end{aligned}
\end{equation}
where the enthalpy $w$ is given by
\begin{equation}
    w = \frac{\pi^2 n^2}{2}-\frac{1}{4} \partial_z^2 \ln n - \frac{1}{8}\left(\partial_z \ln n\right)^2 + \widehat{A} n \; ,
    \label{eq:enthalpy}
\end{equation}
and the action of the operator $\widehat{A}$ in momentum space is $\widehat{A} n_k=\pi|k| n_k$~\cite{JEVICKI199275}.
The first term of eq.~\eqref{eq:enthalpy} is the pressure of a homogeneous Fermi gas, the two  middle terms arise from gradient corrections and describe the cyclotronic pressure, while the final term accounts for loop corrections.
Interestingly, the contributions of the former effects also appear in the GPE.
However, for free fermions, the analysis of Ref.~\cite{PhysRevB.87.045112} proved that the latter effect is more important.

Applying Whitham's approach to the hydrodynamics above, one obtains the following expression for the shock wave density
\begin{equation}
    n = n_1-\frac{k_{32}}{2 \pi} \frac{\sinh a}{\cosh a-\cos \phi} \; .
    \label{eq:densit_Whitham_FF}
\end{equation}
Here, $\phi = k_{32} z - \omega t$ is the phase of the oscillating wave, $n_1 = n_0 + k_{32} / (2 \pi)$, while the parameter $a$ control the amplitude and shape of the wave and reads
\begin{equation}
    \tanh a=\frac{4 \pi k_{32}^3 \rho_1^3}{k_{32}^4 n_1^2+4 \pi^2 k_{32}^2 n_1^4-4\left(k_{32} j-\omega n_0\right)^2} \; .
\end{equation}
The wave vector of the wave $k_{32}$, its frequency $\omega$, as well as its mean density and current are given by 
\begin{equation}
\begin{aligned}
k_{32} &= \Lambda_3 - \Lambda_2 \; , \\
\omega &= \frac{1}{2}\left(\Lambda_3^2-\Lambda_2^2\right) \; , \\
n_0 &= \frac{\Lambda_4 - \Lambda_3 + \Lambda_2 - \Lambda_1}{2 \pi} \; , \\
j &= \frac{\Lambda_4^2 - \Lambda_3^2 + \Lambda_2^2 - \Lambda_1^2}{4 \pi} \; .
\end{aligned}
\end{equation}
Note, the equations above were originally phrased in terms of the Fermi momenta of the post-shock Wigner function, however, in the Tonks-Girardeau regime, these are equivalent to the Fermi rapidities of GHD. 
Thus, for the sake of consistent notation, we express the equations in terms of the latter.

In addition to the hydrodynamic approach, Ref.~\cite{PhysRevB.87.045112} also conducted a semi-classical analysis using the Wigner function.
Naively, one would expect the Wigner function of a free fermion gas in a potential well to behave similarly to the filling function of the thermodynamics Bethe Ansatz; within the Fermi momenta it assumes the value 1, while outside it abruptly drops to 0.
In fact, the dynamics and resulting density of such a Wigner function are equivalent to the zero temperature GHD of the Tonks-Girardeau gas.
However, in the correct semi-classical approximation of the Wigner function, the function develops oscillations in momentum space near the Fermi momentum.
The number of oscillations is determined by the number of particles in the initial perturbation.
Following dynamics, these oscillations in momentum space translate into density oscillations within the shock wave region.
When compared to "exact" results obtained from direct quantum simulations of free fermions on a lattice, this semi-classical approach reproduces the shock wave oscillations almost faithfully.
However, the oscillations of the quantum simulation are more irregular, particularly around the center of the shock wave region; note that the results of the quantum simulations are correctly captured by QGHD~\cite{private_comm_stefano}.
Nevertheless, the semi-classical Wigner approach will be used as our reference solution in the following discussion.

In figure~\ref{fig:TG_Whitham} we plot the oscillation period obtained from the FF Whitham~\eqref{eq:densit_Whitham_FF}, the semi-classical Wigner function, and the GPE Whitham solution~\eqref{eq:density_Whitham}.
The physical setup in question is a density perturbation travelling on a homogeneous background, where the underlying data is taken from Ref.~\cite{PhysRevB.87.045112}.
Immediately we observe that the GPE Whitham fails to capture the "correct" oscillation period, whereas the FF Whitham and semi-classical Wigner methods converge when approaching the front edge of the shock.
Indeed, the wave vector $k_{32}$ of the FF Whitham coincides with $k_{42} = \Lambda_4 - \Lambda_2$ at the front edge, since $\Lambda_3 \to \Lambda_4$.
Evidently, the oscillations of the Wigner approach are accurately captured by the wave vector $k_{42}$, however, the reason for this is not fully apparent. 
Notably, the oscillation period of neither the FF Whitham nor the semi-classical Wigner approach explicitly depend on the lower Fermi momentum $\Lambda_1$, unlike the GPE oscillations.
As a result, the effective shock wave vector of the GPE solution is much larger (also the oscillating term $\mathrm{sn}(\, \cdot\,)$ is squared).

\section{The bipartition problem with different background densities} \label{app:Riemann_background_variantion}

\begin{figure}
    \centering
    \includegraphics[width = 1\columnwidth]{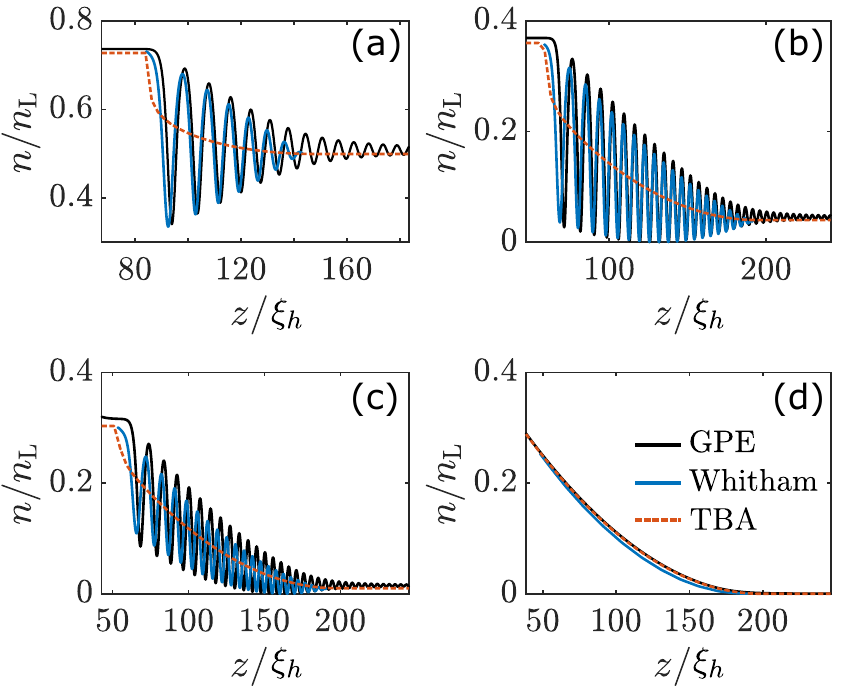}

    \caption{\label{fig:Riemann_variable_density}
    Dispersive shock wave in the bipartition problem at time $t v_s /l = 100$ for various densities on the right boundary; (\textbf{a}) $n_\mathrm{R} = 0.5 n_\mathrm{L}$, (\textbf{b}) $n_\mathrm{R} = 0.04 n_\mathrm{L}$, (\textbf{c}) $n_\mathrm{R} = 0.01 n_\mathrm{L}$, and (\textbf{d}) $n_\mathrm{R} = 0$.
    The density is calculated from the Fermi contours of GHD simulations using the Bethe Ansatz~\eqref{eq:density_expectation_value_filling} and Whitham theory~\eqref{eq:density_Whitham} and compared with results from GPE simulations.
    }
\end{figure}

The characteristics of a DSW depend not only on the traveling wave itself, but also the background on which it propagates; in particular the density of the background, here $n_\mathrm{R}$, is fundamental to the shape of the DSW.
In order to demonstrate this behavior, we repeat the GHD and GPE simulations of section~\ref{sec:zero_temp} for varying densities on the right side of the initial boundary and plot the resulting DSWs after an evolution time $t v_s /\xi_h = 100$ in figure~\ref{fig:Riemann_variable_density}. 
We find that for small values of $n_\mathrm{R}$, the amplitude of the shock wave oscillations decreases until the DSW fully disappears for $n_\mathrm{R} = 0$.
Such behavior is well-known in the bipartition problem~\cite{EL1995186}; consider the soliton on the border with the central density plateau, whose amplitude is
\begin{equation}
    A = 2 (\sqrt{n_\mathrm{L} n_\mathrm{R}} - n_\mathrm{R} ) \; .
\end{equation}
Starting with equal density on either side of the initial boundary $n_\mathrm{L} = n_\mathrm{R}$ and gradually decreasing $n_\mathrm{R}$, the difference between $A$ and the plateau density $\bar{n} = \frac{1}{4} \left( \sqrt{n_\mathrm{L}} + \sqrt{n_\mathrm{R}} \right)^2 $ decreases.
For the densities $n_\mathrm{R} = n_\mathrm{L} /9$, we obtain $A = \bar{n}$ and the first soliton becomes black as the condensate density at its minimum reaches zero.
As $n_\mathrm{R}$ further decreases, the amplitude of oscillations in the DSW decreases, while the point of zero density moves deeper into the DSW.
Finally, as the density on the right side vanishes, the entire traveling wave solution becomes a rarefaction wave.
The results of figure~\ref{fig:Riemann_variable_density}  reflect exactly this behavior; comparing the oscillating solutions of the GPE simulations with the DSW density obtained from the GHD simulations via eq.~\eqref{eq:density_Whitham}, we again find a good agreement.
This demonstrates a certain robustness of our approach while providing further example for the physical interpretation of DSW formation as the interference of the two Fermi seas.

\section{The constantly accelerating piston} \label{sec:piston_problem}
\begin{figure*}
    \centering
    \includegraphics[width = 1\textwidth]{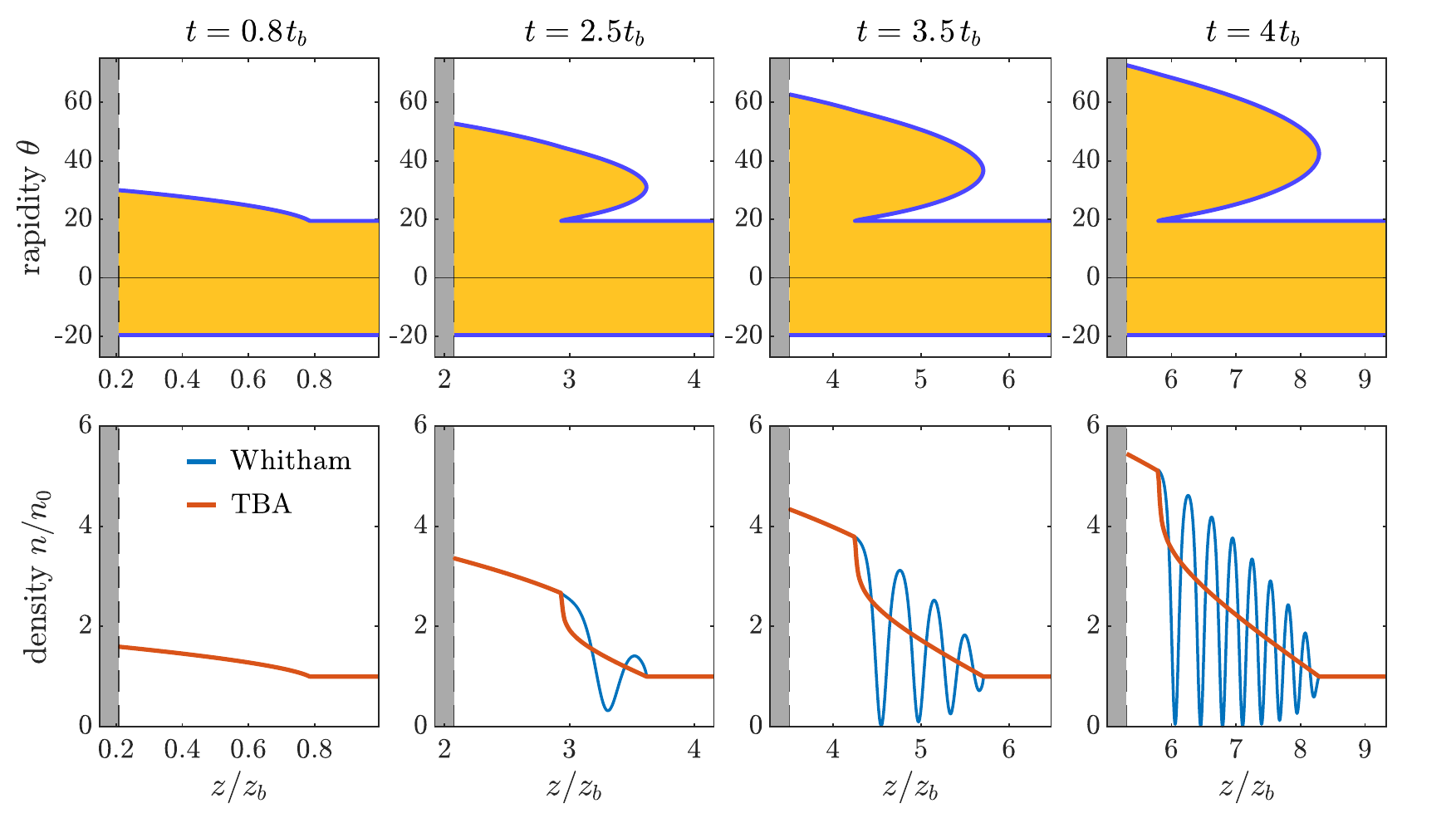}

    \caption{\label{fig:piston_snapshots}
    Simulated Fermi contours $\Gamma_\pm$ of a condensate under the action of a constantly accelerating piston plotted at select times following the breaking time $t_b$.
    Below, the corresponding density profiles computed using Thermodynamic Bethe Ansatz (eq.~\eqref{eq:density_expectation_value_filling}, plotted in red) and Whitham's theory (eq.~\eqref{eq:density_Whitham}, plotted in blue) are shown.
    The grey shaded area marks the position of the piston.
    }
\end{figure*}

\begin{figure}
    \centering
    \includegraphics[width = 1\columnwidth]{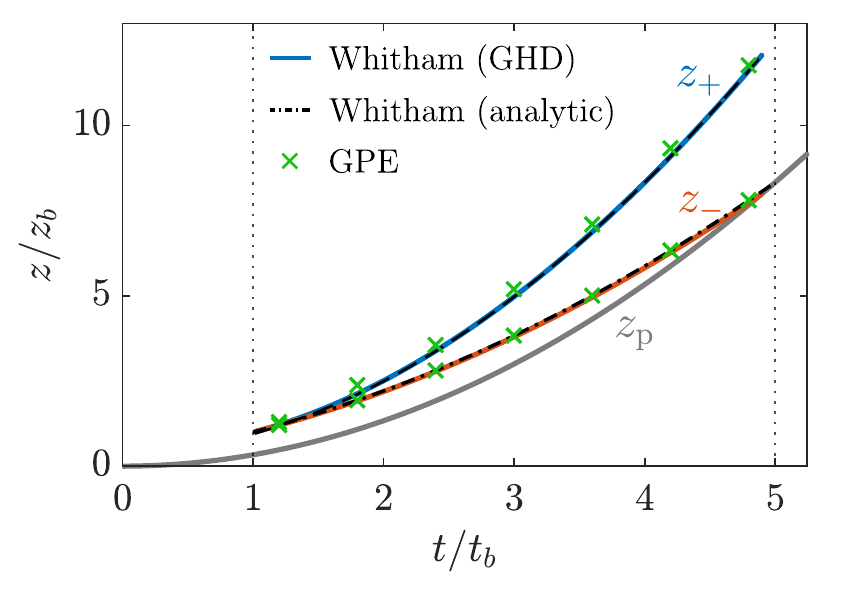}

    \caption{\label{fig:piston_DSW_edges}
    Position of the dispersive shock wave boundaries over time.
    The results of GHD simulations are represented as colored lines; the black dashed and dotted lines denote the analytic predictions of eqs.~\eqref{eq:DSW_piston_trailing_edge} and \eqref{eq:DSW_piston_leading_edge}; the green crosses are the results of GPE simulations.
    The breaking time $t_b$ of eq.~\eqref{eq:piston_breaking_time} heralds the formation of the DSW, while the critical time $t_c$ of eq.~\eqref{eq:piston_critical_time} marks the point at which the piston boundary, plotted as a solid grey line, reaching the edge of the shock wave. 
    }
\end{figure}

Lastly, we consider the action of a piston moving through a homogeneous condensate with density $n_0$ initially at rest.
To this end, we consider a constantly accelerating piston, whose position and velocity are denoted $z_p (t) = a_p t^2 /2$ and $v_p (t) = a_p t$, where $ a_p > 0$ and is constant.
This problem was first treated using Whitham's approach in Ref.~\cite{Kamchatnov2010}; here we summarize the main analytical predictions, which we will later compare with the results of GHD simulations.

The piston acts only at its boundary with the condensate, whereby its effect is taken into account through the boundary condition 
\begin{equation}
    u(z,t) = v_p(t) \quad \text{for } z = z_p(t) \; ,
    \label{eq:piston_boundary}
\end{equation}
which specifies that the hydrodynamic velocity of the condensate must be equal to the piston velocity at their point of contact~\cite{PhysRevLett.100.084504}. 
At sufficiently short times, the piston velocity is lower than the sound velocity of the background condensate $v_{s0} = \hbar \sqrt{n_0 c} /m$, whereby the wave does not break immediately as in the bipartition problem. 
Instead, similarly to the illustration of figure~\ref{fig:illustration}, the breaking time $t_b$ denotes the time at which a vertical tangent appears in the profiles of the density and hydrodynamic velocity, corresponding to the development of the second Fermi sea in GHD.
Before the breaking time, the condensate consists of two distinct regions: In the region $z_p(t) < z < v_{s0} t$ the hydrodynamic velocity is positive as the condensate has been accelerated by the piston, while for $z > v_{s0} t$ the condensate remains at rest.
At the boundary between the two regions, the following condition must be met
\begin{equation}
    n = n_0\;, \quad u = 0 \quad \text{for } z = v_{s0} t \; .
\end{equation}
In the bulk of the condensate the gradient of the density is small, whereby the flow can be described by the expressions of the dispersion-less regime.
From these considerations one finds the breaking time to be 
\begin{equation}
    t_b = \frac{2 v_{s0}}{3 a_p} \; ,
    \label{eq:piston_breaking_time}
\end{equation}
and the point of the initial wave breaking to be 
\begin{equation}
    z_{b} = \frac{2 v_{s0}^2}{3 a_p} \; .
\end{equation}
When approaching the breaking point, dispersion becomes relevant and a DSW starts forming near $z_b$. 
By considering the boundary conditions of the Whitham equations at the DSW edges, it is possible to obtain a complete description of the condensate flow under the action of the piston~\cite{Kamchatnov2010}.
Thus, the position of the trailing (soliton) edge of the DSW is given by
\begin{equation}
    z_{-} (t) = \frac{5}{36} \frac{v_{s0}^2}{a_p} + \frac{7}{12} v_{s0} t + \frac{5}{16} a_p t^2 \; ,
    \label{eq:DSW_piston_trailing_edge}
\end{equation}
while the position of the leading (low amplitude) edge reads
\begin{equation}
    z_{+} (t) = \frac{2 v_{s0}^2}{5 a_p} \frac{8 y^4 - 6y^2 + 3}{2 y^2 + 1} \; ,
    \label{eq:DSW_piston_leading_edge}
\end{equation}
where $y = \mathrm{min} (r_4) /v_{s0}$. 
Finally, at the critical time  
\begin{equation}
    t_c = \frac{10 v_{s0}}{3 a_p} \; ,
    \label{eq:piston_critical_time}
\end{equation}
the piston catches up to the trailing edge of the DSW, whereby an additional (third) Fermi sea will develop. 
Analysis of the piston problem past this point is beyond the scope of this work.

In order to simulate the piston problem using zero-temperature GHD, we proceed in a similar manner to the bipartition problem by discretizing the Fermi contour as a number of points.
Given the piston boundary condition~\eqref{eq:piston_boundary}, the Riemann invariants take the form
\begin{equation}
    r_{+} - v_{s0} = v_p \quad \text{for } z =z_p(t) \; ,
\end{equation}
while $r_{-} = - v_{s0}$ remains a constant invariant.
Thus, the action of the piston can be seen as reflection of the lower Riemann invariant plus a local boost by the piston velocity.
For the GHD simulation, we assume that the piston acts on the Fermi contours in an equivalent manner.

In figure~\ref{fig:piston_snapshots} we plot the Fermi contour at a few evolution times $t$ simulated for $n_0 = 200$, $c = 0.5$, and $a = 0.1 v_{s0}^2 / \xi_h$.
By virtue of the accelerating piston, the width of the Fermi sea at the solition edge of the DSW grows increasingly wider, leading to an increasing density difference across the DSW. 
Furthermore, the rapidity at the leading edge of the DSW $z_{+}$ increases, indicating that the boundary is accelerating.
This in contrast to the constant velocity piston~\cite{PhysRevLett.100.084504}, where both the density difference and the boundary velocity remain constant.
Once again we compute the periodic density solution from the Fermi contour and plot the results in figure~\ref{fig:piston_snapshots}.
As the width of the second Fermi sea grows, the amplitude of the DSW oscillations increases; as evident from eq.~\eqref{eq:density_Whitham}, once the width of the Fermi sea corresponding to the traveling wave equals that of the background, the first soliton of the shock wave train becomes black.
Increasing the width of the second Fermi sea beyond this point further increases the number of black solitons.
Once again we compute the GHD density and find that it accurately represents the average density of the DSW.

Next, we extract the positions on the Fermi contour, where the state transitions from one to two Fermi seas, and plot the results in figure~\ref{fig:piston_DSW_edges} along with the position of the piston.
At time $t_b$ the wave breaks and the DSW forms; by virtue of their different velocities, the width of the DSW increases over time until the critical point $t_c$, where the piston reaches the trailing edge of the wave.
Comparing the results of the GHD simulation to the analytic expressions for the trailing and leading edge positions of the DSW given by eqs. \eqref{eq:DSW_piston_trailing_edge} and \eqref{eq:DSW_piston_leading_edge}, respectively, we find a very good agreement. 
Upon careful inspection a small discrepancy can be found, which we attribute to the finite coupling strength employed in the GHD simulation.
For the sake of completeness, we also compare our results with GPE simulations and find a good agreement, as seen in figure~\ref{fig:piston_DSW_edges}.

\bibliography{references}

\end{document}